\documentclass[graybox]{svmult}

\usepackage{amsmath}
\usepackage{amssymb}
\usepackage{mathptmx}       
\usepackage{helvet}         
\usepackage{courier}        
\usepackage{type1cm}        
\usepackage{makeidx}         
\usepackage{graphicx}        
\usepackage{multicol}        
\usepackage[bottom]{footmisc}

\makeindex             

\begin{document}

\title*{Limit Distribution of Averages over Unstable Periodic Orbits
Forming Chaotic Attractor}
\titlerunning{Limit Distribution of Averages over UPOs in Chaotic Attractors}
\author{Denis S.\ Goldobin}
\authorrunning{D.S.\ Goldobin}
\institute{Denis S.\ Goldobin \at Institute of Continuous Media Mechanics, UB RAS,
             1 Acad.\ Korolev str., Perm 614013, Russia
           \and Department of Theoretical Physics, Perm State University,
             15 Bukireva str., Perm 614990, Russia
              \email{Denis.Goldobin@gmail.com}}
%
%
\maketitle

\abstract*{We address the question of representativeness of a single long
unstable periodic orbit for properties of the chaotic attractor it
is embedded in. Y. Saiki and M. Yamada [Phys. Rev. E 79, 015201(R)
(2009)] have recently suggested the hypothesis that there exist a
limit distribution of averages over unstable periodic orbits with
given number of loops, N, which is not a Dirac delta-function for
infinitely long orbits. In this paper we show that the limit
distribution is actually a delta-function and standard deviations
decay as 1/sqrt(N) for large enough N.}

\abstract{We address the question of representativeness of a single long
unstable periodic orbit for properties of the chaotic attractor it
is embedded in. Y.~Saiki and M.~Yamada [Phys.\,Rev.\,E {\bf 79},
015201(R) (2009)] have recently suggested the hypothesis that
there exist a limit distribution of averages over unstable
periodic orbits with given number of loops, $N$, which is not a
Dirac $\delta$-function for infinitely long orbits. In this paper
we show that the limit distribution is actually a
$\delta$-function and standard deviations decay as $N^{-1/2}$ for
large enough $N$.}

\section{Introduction}
Recent investigations~\cite{Saiki-Yamada-2009, Kawahara-Kida-2001,
Kato-Yamada-2003, vanVeen-Kidaa-Kawahara-2006, Nikitin-2007} have
arose the question of representativeness of a single unstable
periodic orbit (UPO) for properties of the chaotic attractor it is
embedded in. Interestingly, while Refs.~\cite{Kawahara-Kida-2001,
Kato-Yamada-2003, vanVeen-Kidaa-Kawahara-2006} concern unstable
time-periodic orbits in turbulence, Ref.~\cite{Nikitin-2007}
discusses turbulent pipe flows in relation to spatially periodic
solutions. Motivated by~\cite{Kawahara-Kida-2001,
Kato-Yamada-2003, vanVeen-Kidaa-Kawahara-2006, Nikitin-2007}
Y.~Saiki and M.~Yamada~\cite{Saiki-Yamada-2009} discus the
calculation of average values along UPOs (unstable periodic orbits
embedded into the chaotic attractor) and using the results of such
a calculation for estimation of averages along a chaotic
trajectory. In particular, with simple paradigmatic chaotic
systems---the Lorenz system, the R\"ossler one, and a
6-dimensional economic model---Y.~Saiki and M.~Yamada explored the
convergence of average values calculated along UPOs as the number
of loops, $N$, of a UPO grows. They surprisingly found that the
distribution of average values along all the UPOs with given $N$
(henceforth, $N$-UPOs) practically does not shrink for long orbits
($N=10...14$ for the Lorenz system, $N=10...17$ for the R\"ossler
one), suggesting the existence of a limit distribution of averages
over $N$-UPOs, which is not a Dirac $\delta$-function for
infinitely long orbits ($N\to\infty$). In
Ref.~\cite{Zaks-Goldobin-2010}, it was shown that the UPOs with
the length $N\le14$ are not enough long for conclusions on
asymptotic behavior.

We are interested in the asymptotic behavior of the distribution
of average values over UPOs with fixed $N$. Notice, the evaluation
of average values for the chaotic attractor requires this
distribution to be weighted with the distribution of the natural
measure over the set of UPOs; in
Ref.~\cite{Grebogi-Ott-Yorke-1988} the natural measure of an UPO
was shown to be inverse proportional to its multiplier. Meanwhile,
we restrict our consideration to the question of convergence of
the limit distribution and do not consider the distribution of the
natural measure.

\begin{figure}[t]
\center{
 (a)\hspace{-10pt}
 \includegraphics[width=0.26\textwidth]{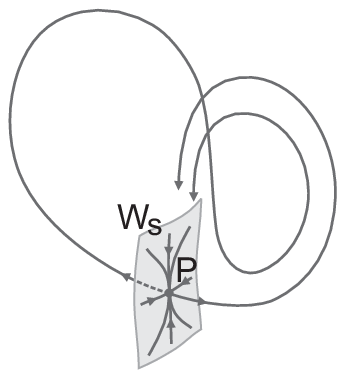}\qquad
 \includegraphics[width=0.42\textwidth]{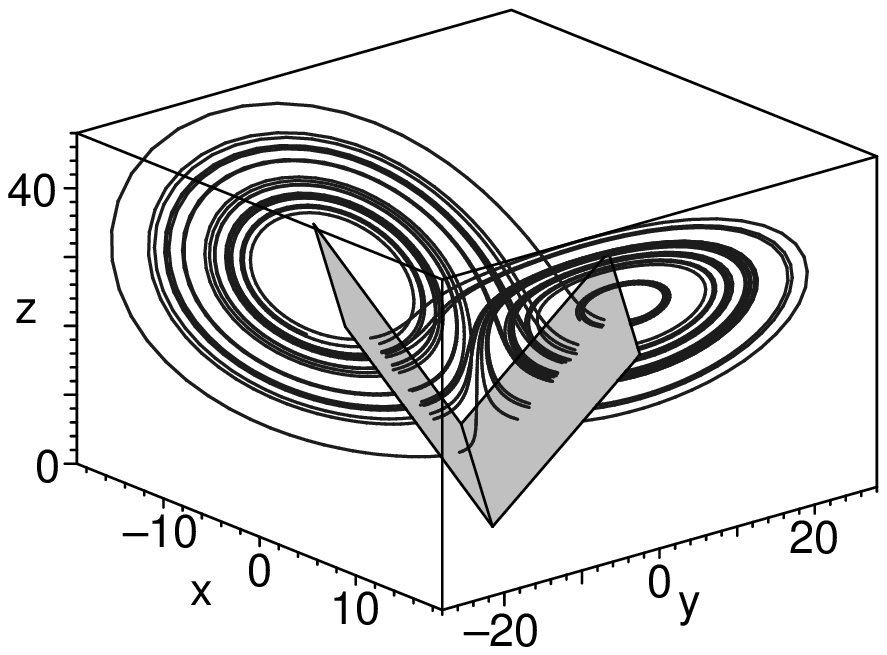}\\[15pt]
 (b)\hspace{-10pt}
 \includegraphics[width=0.30\textwidth]{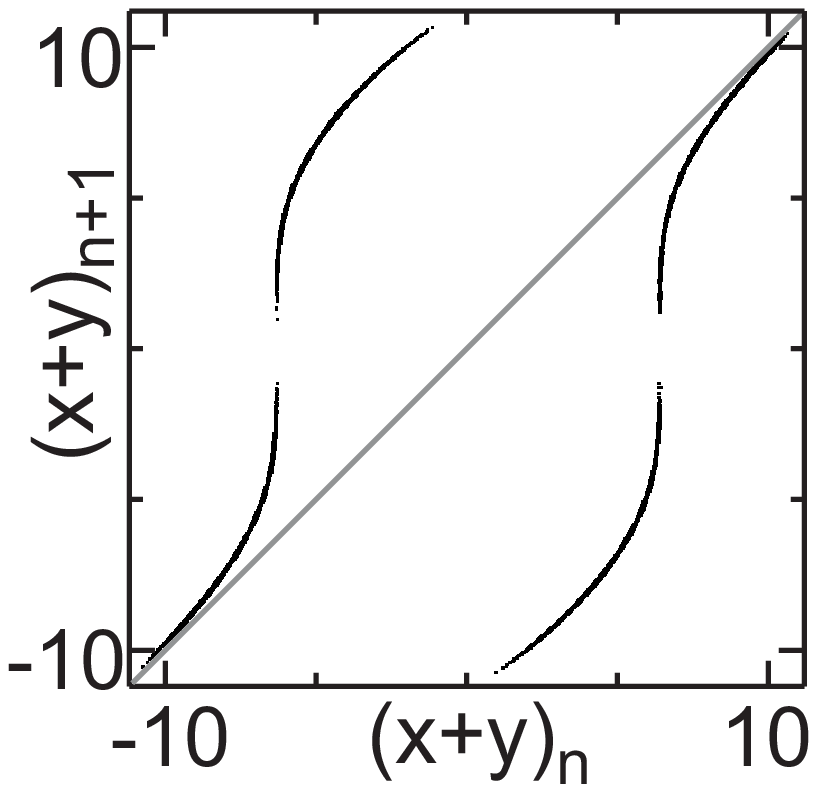}\qquad\qquad
 (c)\hspace{-10pt}
 \includegraphics[width=0.30\textwidth]{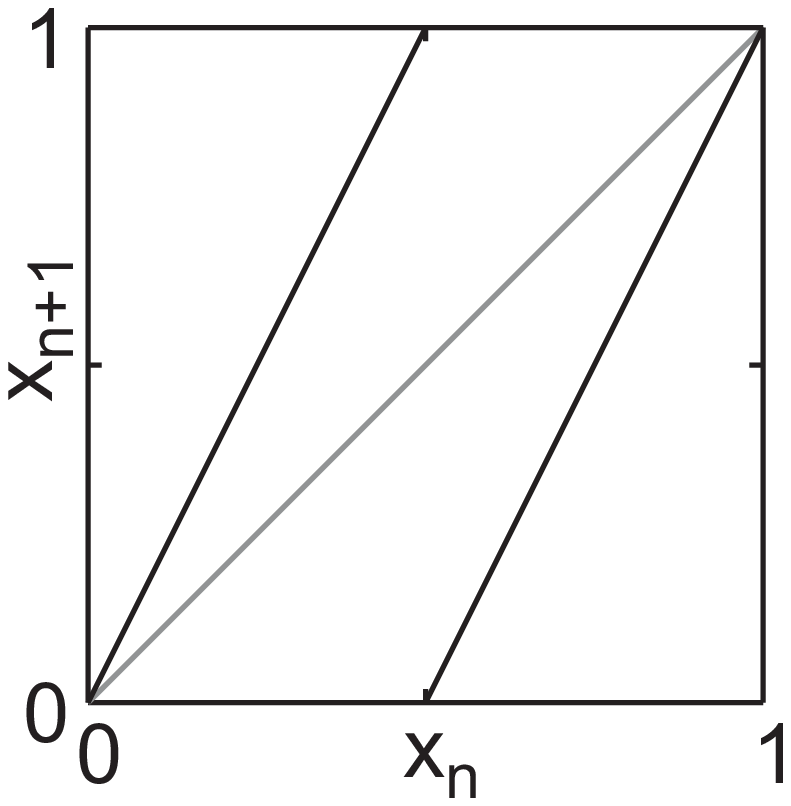}
 }
\caption{Chaos arising via bifurcations of homoclinic orbits.
The phase portrait and the Poincar\'e section for the Lorenz
system with classical parameter values (a). The Poincar\'e map for
$(x+y)$ (b) is qualitatively similar to the saw map (c).}
\label{fig1}
\end{figure}

\begin{figure}[t]
  \center{
 (a)\hspace{-10pt}
 \includegraphics[width=0.42\textwidth]{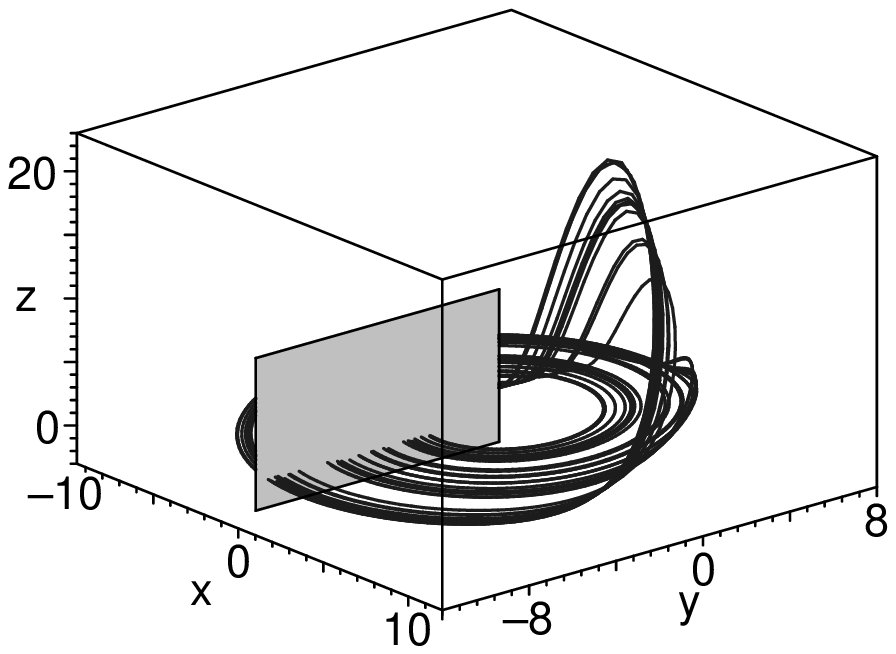}
 \qquad\qquad
 (b)\hspace{-10pt}
 \includegraphics[width=0.30\textwidth]{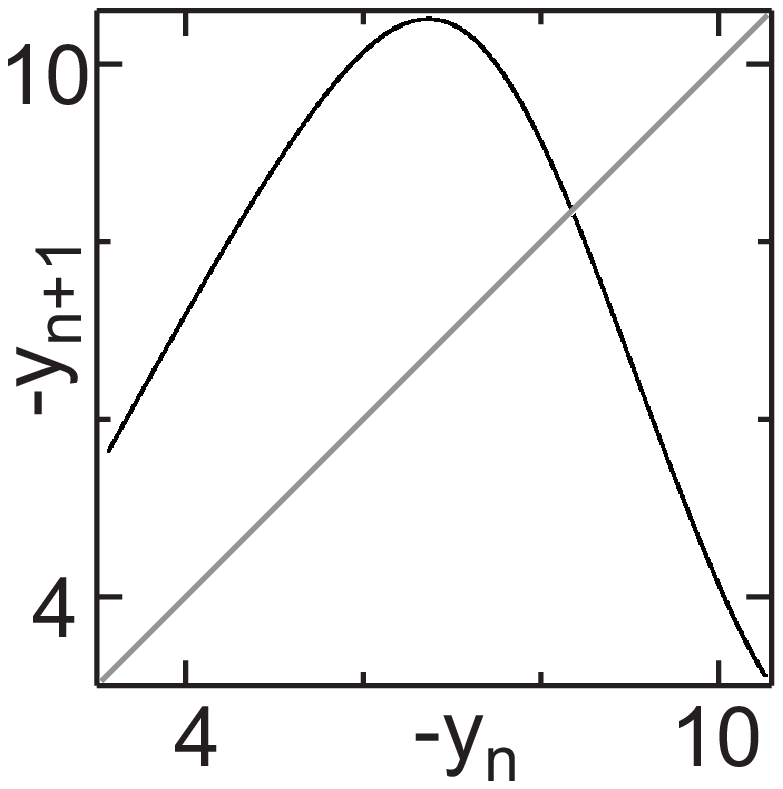}
 }
  \caption{Chaos arising via cascade of the period-doubling bifurcations.
The Poincar\'e section (a) and map (b) for the R\"ossler system
with classical parameter values .}
  \label{fig2}
\end{figure}

In this work we treat two maps: ``saw'' map
\begin{equation}
x_{n+1}=2x_n\mod{1}\,
\label{saw-m}
\end{equation}
---paradigmatic model
for the chaos (i) arising via cascade of bifurcations of
homoclinic orbits (as in the Lorenz system~\cite{Lorenz-1963},
Fig.~\ref{fig1}a)~\cite{Kaplan-Yorke-1979,
Lyubimov-Zaks-1983}---and ``tent'' map
\begin{equation}
x_{n+1}=\left\{\begin{array}{cc}
 2x_n& \mbox{for }x_n\le1/2\,,\\[10pt]
 2-2x_n& \mbox{for }x_n>1/2\,
\end{array}\right.
\label{tent-m}
\end{equation}
---paradigmatic model
for the chaos (ii) arising via cascade of the period-doubling
bifurcations (as in the R\"ossler system,
Fig.~\ref{fig2}a)~\cite{Feigenbaum-1978,Feigenbaum-1979,Collet-Eckmann-Koch-1981}.
The reason for our choice is not only paradigmaticity of these
maps but also the fact that one gains exceptional technical
opportunities for calculating all UPOs in them for giant $N$. We
find the distributions of averages upon $N$-UPOs to shrink as $N$
grows; standard deviations decay as $1/\sqrt{N}$. Then we consider
the Lorenz (the first sort of chaotic systems) and find for it the
same kind dependence of standard deviations on $N$ as for the
paradigmatic map models. Convergence of the distribution to a
$\delta$-function can be observed for enough long UPOs (for the
Lorenz system $N\ge10$, cf.\ Fig.~\ref{fig4}).

\begin{figure}[t]
\center{
 (a)\hspace{-15pt}
 \includegraphics[width=0.42\textwidth]{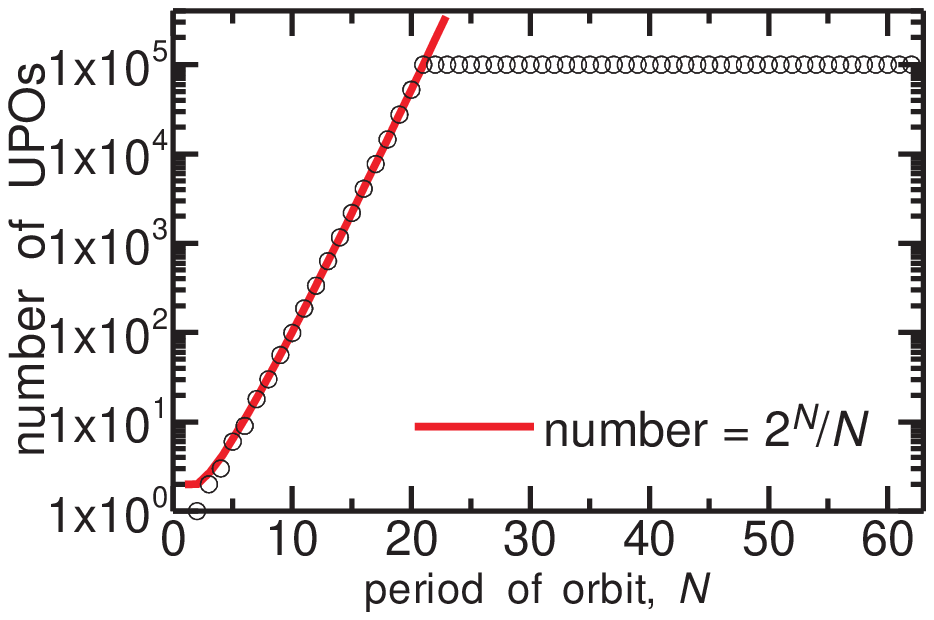}\qquad\qquad
 (b)\hspace{-15pt}
 \includegraphics[width=0.42\textwidth]{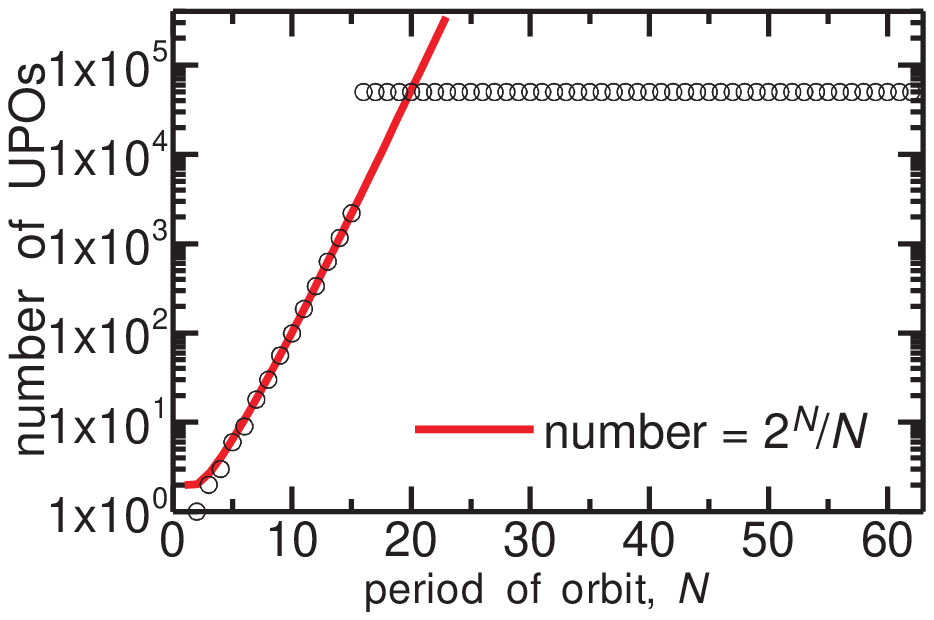}\\[15pt]
 (c)\hspace{-15pt}
 \includegraphics[width=0.42\textwidth]{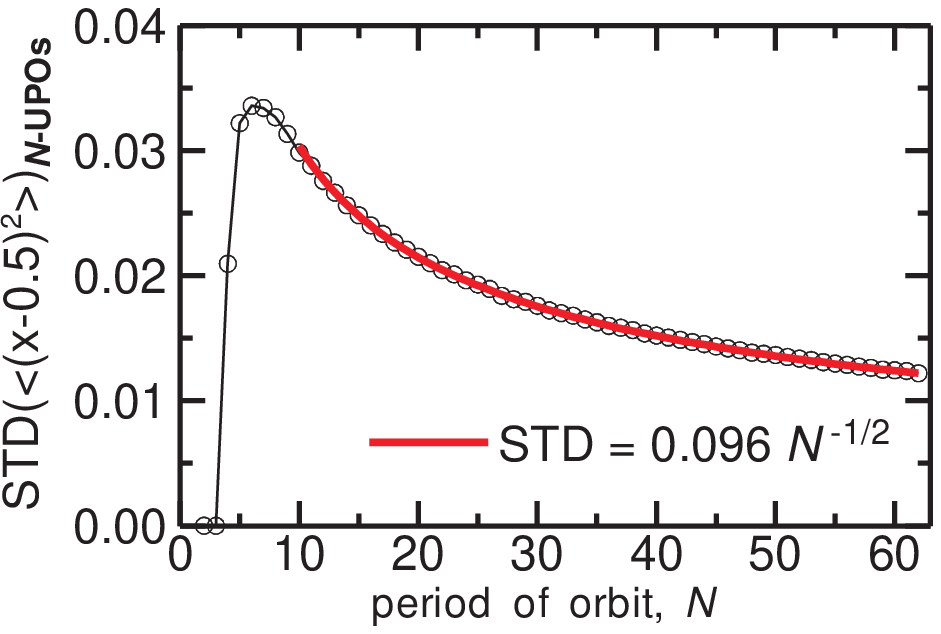}\qquad\qquad
 (d)\hspace{-15pt}
 \includegraphics[width=0.42\textwidth]{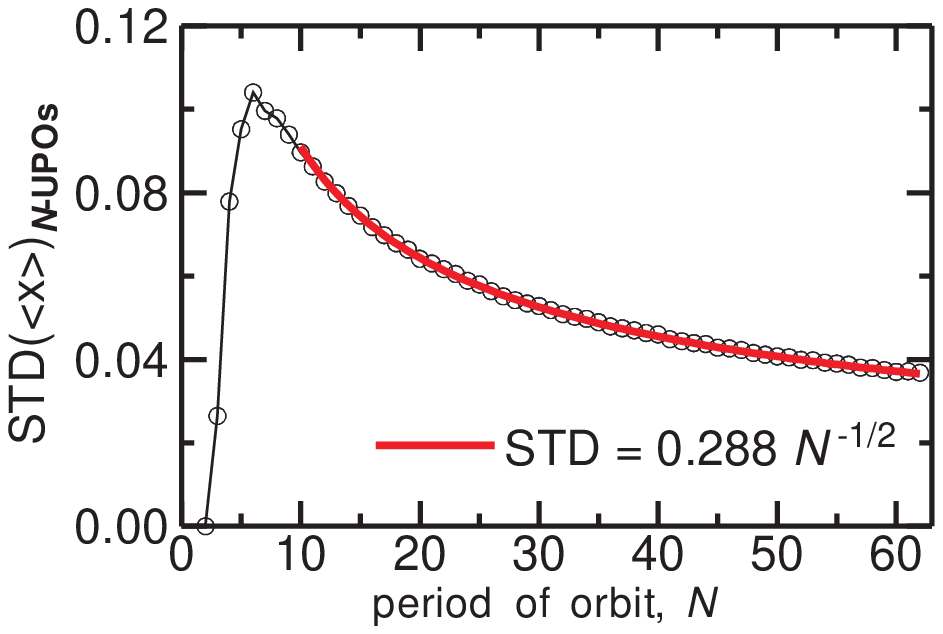}\\[15pt]
 (e)\hspace{-15pt}
 \includegraphics[width=0.42\textwidth]{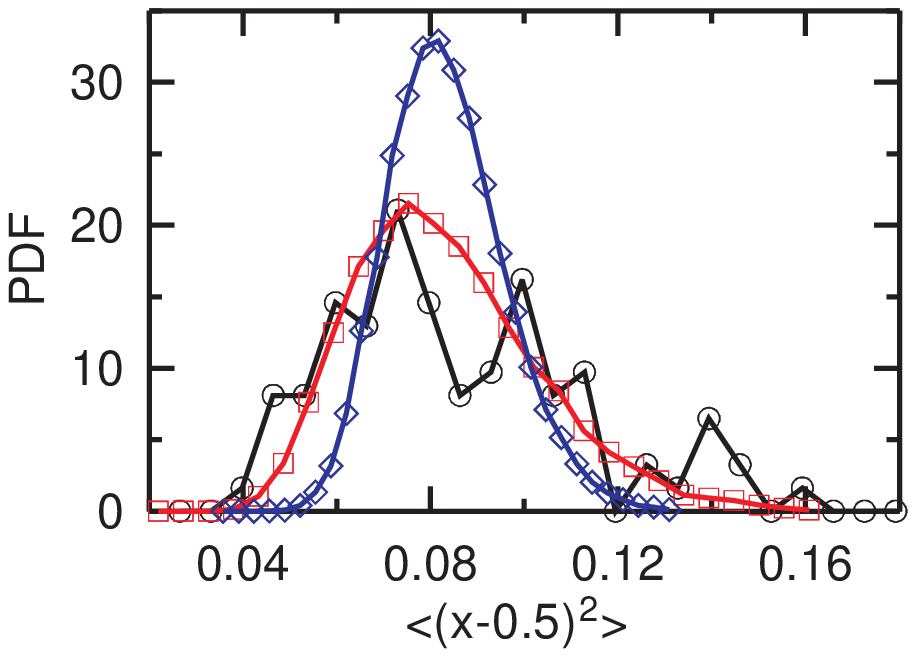}\qquad\qquad
 (f)\hspace{-15pt}
 \includegraphics[width=0.42\textwidth]{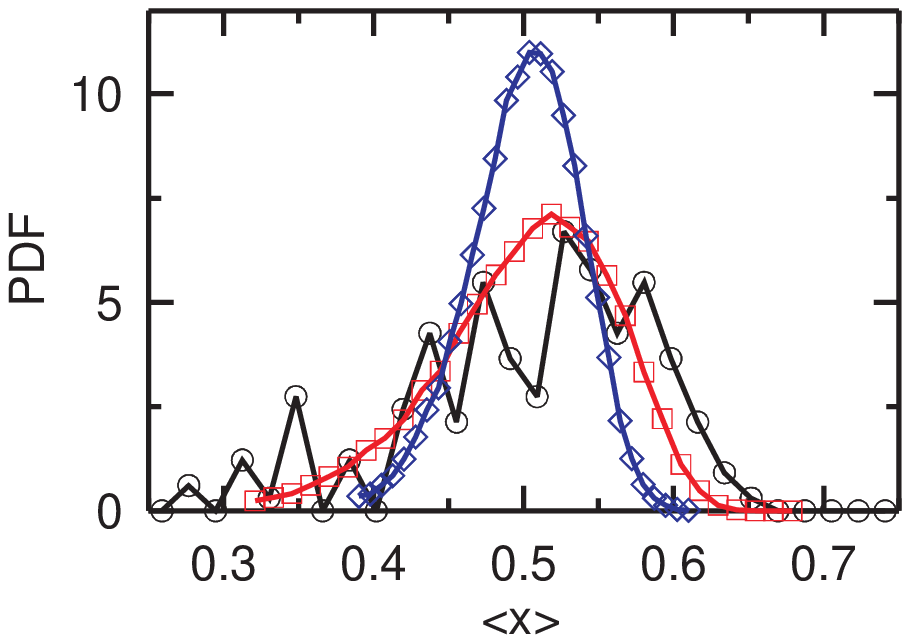}\\
 }
  \caption{The number of UPOs used for averaging
is presented by circles for the saw (a) and the tent (b) map;
until the plateau, all UPOs are detected and employed, and on the
plateau we utilize UPOs chosen randomly with a probability uniform
over all $N$-UPOs. The standard deviations of
$\langle(x-1/2)^2\rangle$ for the saw map (c) and the one of
$\langle{x}\rangle$ for the tent map (d) decay as $1/\sqrt{N}$ for
enough large $N$. The PDFs of the averages can be seen to shrink
slowly as $N$ grows for both the saw (e) and the tent (f) maps;
circles: $N=11$, squares: $N=23$, diamonds: $N=61$.}
  \label{fig3}
\end{figure}

\section{Limit Distribution of Averages}
Let us first recall the origin of maps (\ref{saw-m}) and
(\ref{tent-m}) in order to make the physical representativeness of
these models evident. In the systems where the both separatrices
of the saddle knot ${\bf P}$ come back in the vicinity of the
stable manifold $W_s$ of ${\bf P}$ (Fig.~\ref{fig1}a) the touching
of a separatrix and $W_s$ results in appearance of the homoclinic
orbit. The cascade of bifurcations of such orbits leads to the
formation of the chaotic set (in the Lorenz system this cascade
occurs at the single point $r\approx13.927$ owing to the symmetry
$(x,y)\leftrightarrow(-x,-y)$, cf.~\cite{Lorenz-1963,
Kaplan-Yorke-1979, Lyubimov-Zaks-1983}). For the Lorenz system
\begin{eqnarray}
&&\frac{dx}{dt}=\sigma(y-x)\,,
\nonumber\\
&&\frac{dy}{dt}=rx-y-xz\,,
\label{Lorenz}\\
&&\frac{dz}{dt}=-bz+xy
\nonumber
\end{eqnarray}
with the classical parameter set [$\sigma=10$, $b=8/3$, $r=28$] we
choose $z=[(r-1)/(4b)]^{1/2}|x+y|$ as a Poincar\'e section, and
find that the transversal structure of its intersection with the
chaotic set is very narrow (Fig.~\ref{fig1}); the Poincer\'e map
for $(x+y)$ is visually unambiguous (Fig.~\ref{fig1}b) and similar
to the saw map (Fig.~\ref{fig1}c). The averaging of a certain
value along a certain trajectory on the chaotic set can be reduced
to the averaging of a certain function $f(x+y)$ along the
trajectory in the Poincar\'e map of $(x+y)$. The system symmetry
suggests an even function $f(x+y)=f(-x-y)$. Actually, the
contribution of $f(x+y)$ into $\langle{f(x+y)}\rangle$ over an
orbit after one Poincar\'e recurrence  is weighted by
$T(x+y)/\langle{T(x+y)}\rangle_\mathrm{map}$, the ratio of the
recurrence time for the trajectory running from $(x+y)$ and the
average recurrence time for the orbit,
$\langle...\rangle_\mathrm{map}$ stands for averaging over
iterations of the Poincar\'e map (not over real time). Hence,
 $\langle{f(x+y)}\rangle=
 \langle{f(x+y)\,T(x+y)}\rangle_\mathrm{map}/\langle{T(x+y)}\rangle_\mathrm{map}$,
{\it i.e.}, $\langle{f(x+y)}\rangle$ is subject to the additional
dispersion due to nonuniformity of
$\langle{T(x+y)}\rangle_\mathrm{map}$ over various orbits.
However, the nature of the dispersion of
$\langle{T}\rangle_\mathrm{map}$ is the same as the one of
$\langle{f\,T}\rangle_\mathrm{map}$, and in this paper we do not
introduce any additional complication into our averagings with the
paradigmatic map models. On the whole, the possibility to
construct the Poincar\'e map in the vicinity of the saddle knot
qualitatively similar to the saw map is the common peculiarity of
the systems with a stable chaotic set of the kind we consider.

With regard to the second sort of chaotic systems, the existence
of a unimodal map is a key feature needed for the cascade of the
period-doubling bifurcations to occur resulting in chaotic
behavior of the
system~\cite{Feigenbaum-1978,Feigenbaum-1979,Collet-Eckmann-Koch-1981}.
Thus, for instance, the Poincar\'e map of the R\"ossler system
(Fig.~\ref{fig2}) is quite similar to parabola of the logistic map
$x_{n+1}=bx_n(1-x_n)$. The attracting chaotic set of the logistic
map possesses its largest size at $b=4$ where this map can be
turned into the tent one (\ref{tent-m}) by virtue of substitution
$x\to\sin{\pi x}$. Hence, the tent map is quite representative for
the systems where chaos arises via the cascade of the
period-doubling bifurcations.

As it is noticed above, the symmetry of the Lorenz system suggests
us consideration of $\langle{f(x)}\rangle$: $f(1/2+x)=f(1/2-x)$
along trajectories in the saw map. Hence, for simplicity, we
consider $\langle(x-1/2)^2\rangle$ for the saw map~(\ref{saw-m})
and $\langle{x}\rangle$ for the tent one~(\ref{saw-m}). For the
Lorenz system we calculate $\langle{x^2}\rangle$ and
$\langle{z}\rangle$.

The dynamics of the saw map~(\ref{saw-m}) can be easily dealt with
within the frameworks of the binary notation of $x$. In this
notation, an iteration of map~(\ref{saw-m}) results in the shift
of the binary point in $x$ by one position to the right and
omitting the integer part. Thus, $x$ with an $N$-periodic sequence
of digits in the binary mantissa belongs to an $N$-periodic orbit.
Employing this fact one can strictly calculate giant amount of
UPOs and averages along them. Dealing in such a fashion with the
tent map~(\ref{tent-m}) is less efficient and more sophisticated,
but still possible. In Fig.~\ref{fig3} one can see that for enough
large $N$ the standard deviations of the averages decay as
$1/\sqrt{N}$. Additionally, for $\langle{x}\rangle$ in the saw
map, which is actually beyond our immediate interest, one can
analytically find probability
 $P(\langle{x}\rangle=m/N)\approx
 N!/[2^N m!(N-m)!]$, where $m=1,2,...,N-1$, and, for $N\gg1$,
 $P(\langle{x}\rangle)\approx (2\pi N)^{-1/2}\exp(-2N\langle{x}\rangle^2)$
and
 $\mathrm{STD}(\langle{x}\rangle)_{N-\mathrm{UPOs}}=(4N)^{-1/2}+O(N^{-1})$.

\begin{figure}[t]
  \center{
 (a)\hspace{-15pt}
 \includegraphics[width=0.420\textwidth]{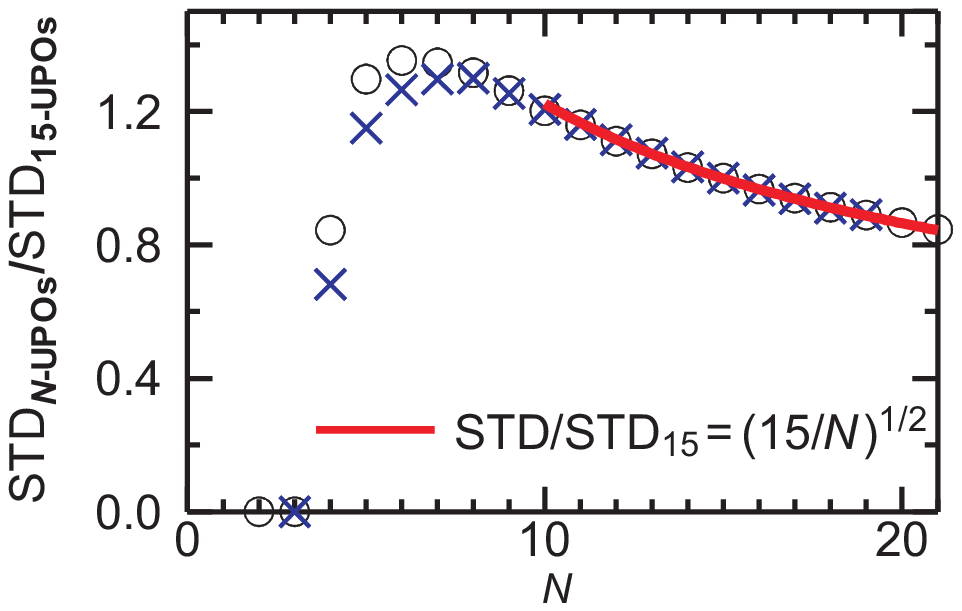}\qquad\qquad
 (b)\hspace{-15pt}
 \includegraphics[width=0.385\textwidth]{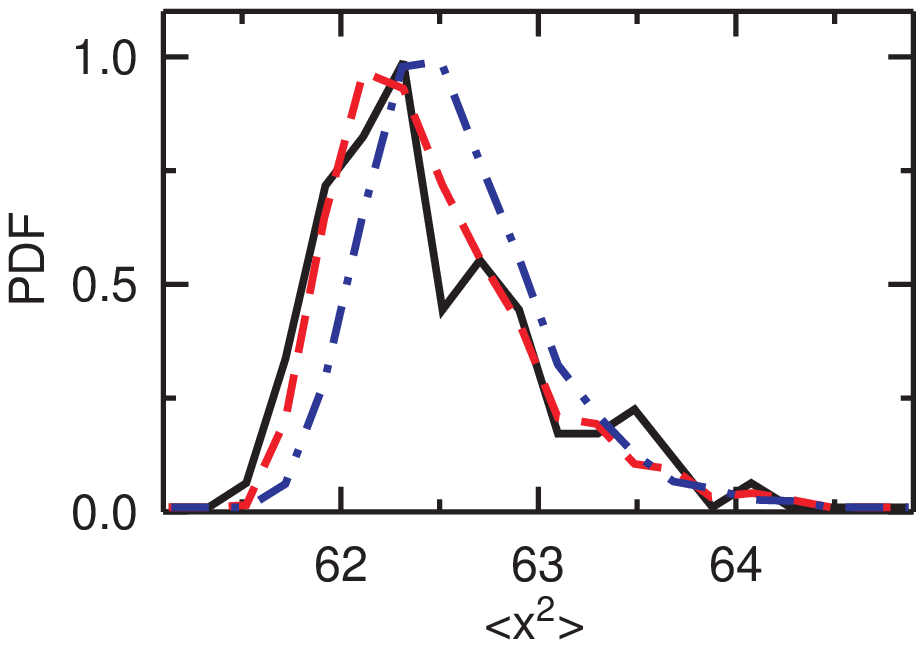}
 }

  \caption{(a): STD normalized by STD for $N=15$ is
plotted by circles for the saw map and by crosses for the Lorenz
system
[\,$\mathrm{STD}(\langle{x^2}\rangle)_{15-\mathrm{UPOs}}\approx0.5086323$\,]
(b): The PDF of $\langle{x^2}\rangle$ for the Lorenz system
shrinks only slightly as $N$ grows at the interval $[10,20]$.
Solid line: $N=11$, dashed line: $N=15$, dash-dotted line:
$N=19$.}
  \label{fig4}
\end{figure}

In order to verify how far our findings for the paradigmatic map
models are relevant for the chaotic systems with continuous time,
UPOs in the Lorenz system~(\ref{Lorenz}) have been numerically
calculated with double precision. We have calculated all UPOs with
$N\le19$ (in Ref.~\cite{Saiki-Yamada-2009} $90\%$ of UPOs with
$N\le14$ were detected) and evaluated $\langle{x^2}\rangle$ and
$\langle{z}\rangle$ along the detected UPOs. In Fig.~\ref{fig4}
one can see the same scaling law for long UPOs;
$\mathrm{STD}\propto N^{-1/2}$. The dependence of STD on $N$ for
the Lorenz system and the saw map are similar for small $N$ as
well (Fig.~\ref{fig4}a).

Notice, the tendency of the PDF to shrinking as $N$ grows can be
hardly detected from plots of PDFs (see Figs.~\ref{fig3}e,f
and~\ref{fig4}b) for $N\in[10,20]$ which were typical for
Ref.~\cite{Saiki-Yamada-2009}.

\section{Conclusion}
Summarizing, we have analyzed two paradigmatic map models: the saw
map~(\ref{saw-m}) representing chaos arising via cascade of
bifurcations of homoclinic orbits and the tent one~(\ref{tent-m})
representing chaos arising via cascade of the period-doubling
bifurcations. For the both models we have reliably established the
fact that the distributions of averages along unstable periodic
orbits with given number of loops $N$ shrinks to a Dirac
$\delta$-function for $N\to\infty$; the standard deviations obey
the decay law $N^{-1/2}$ for $N>10$. For the Lorenz system the
same features have been confirmed. In particular, the average over
a long UPO gives a correct representation of the average over the
whole chaotic set which it is embedded in.

\begin{acknowledgement}
DSG thanks Michael Zaks for fruitful comments on the work. The
work has been financially supported by Grant of The President of
Russian Federation (MK-6932.2012.1). DSG is thankful to Elizaveta
Shklyaeva for motivation to address the subject of this work.
\end{acknowledgement}


\begin{thebibliography}{13.}%

\bibitem{Collet-Eckmann-Koch-1981}
Collet, P., Eckmann, J.-P., Koch, H.:
 Period doubling bifurcations for families of maps on $\mathbb{R}^n$.
 J. Stat. Phys. \textbf{25}, 1--14 (1981).

\bibitem{Feigenbaum-1978}
Feigenbaum, M.J.:
 Quantitative universality for a class of nonlinear
 transformations.
 J. Stat. Phys. \textbf{19}, 25--52 (1978)

\bibitem{Feigenbaum-1979}
Feigenbaum, M.J.:
 The universal metric properties of nonlinear transformations.
 J. Stat. Phys. \textbf{21}, 669--706 (1979)

\bibitem{Grebogi-Ott-Yorke-1988}
Grebogi, C., Ott, E., Yorke, J.A.:
 Unstable Periodic Orbits and the Dimensions of Multifractal Chaotic
 Attractors.
 Phys. Rev. A \textbf{37}, 1711--1724 (1988)

\bibitem{Kaplan-Yorke-1979}
Kaplan, J.L., Yorke, J.A.:
 Preturbulence: A regime observed in a fluid flow model of Lorenz.
 Commun. Math. Phys. \textbf{67}, 93--108 (1979)

\bibitem{Kato-Yamada-2003}
Kato, S., Yamada, M.:
 Unstable periodic solutions embedded in a shell model turbulence.
 Phys. Rev. E \textbf{68}, 025302(R) (2003)

\bibitem{Kawahara-Kida-2001}
Kawahara, G., Kida, S.:
 Periodic motion embedded in plane Couette turbulence: regeneration
 cycle and burst.
 J. Fluid Mech. \textbf{449}, 291--300 (2001)

\bibitem{Lorenz-1963}
Lorenz, E.N.:
 Deterministic Nonperiodic Flow.
 J. Atmos. Sci. \textbf{20}, 130--141 (1963)

\bibitem{Lyubimov-Zaks-1983}
Lyubimov, D.V., Zaks, M.A.:
 Two mechanisms of the transition to chaos in finite-dimensional
 models of convection.
 Physica \textbf{9D}, 52--64 (1983)

\bibitem{Nikitin-2007}
Nikitin, N.:
 Spatial periodicity of spatially evolving turbulent flow caused
 by inflow boundary condition.
 Phys. Fluids \textbf{19}, 091703 (2007)

\bibitem{Saiki-Yamada-2009}
Saiki, Y., Yamada, M.:
 Time-averaged properties of unstable periodic orbits and
 chaotic orbits in ordinary differential equation systems.
 Phys. Rev. E \textbf{79}, 015201(R) (2009)

\bibitem{vanVeen-Kidaa-Kawahara-2006}
van Veen, L., Kidaa, S., Kawahara, G.:
 Periodic motion representing isotropic turbulence.
 Fluid Dyn. Res. \textbf{38}, 19--46 (2006)

\bibitem{Zaks-Goldobin-2010}
Zaks, M.A., Goldobin, D.S.:
 Comment on ``Time-averaged properties of unstable periodic orbits and chaotic orbits in ordinary differential equation
 systems.''
 Phys. Rev. E \textbf{81}, 018201 (2010)

\end{thebibliography}
\end{document}